\documentclass[11pt,a4paper]{article}

\usepackage{units}                                      
\usepackage[T1]{fontenc}
\usepackage{amsfonts, amsmath, amsthm,amssymb,mathtools}
\usepackage{bbm}
\usepackage{oldgerm}                                    
\usepackage{bibgerm}                                    
\usepackage{geometry}
\usepackage{verbatim}

\usepackage[german, english]{babel}

\DeclareMathOperator{\Tr}{Tr}

\title{A proof of the continuous Dyson-Maleev representation}
\author{J. M\"uller-Hill}
\date{December, 21, 2010}
\begin{document}
\newtheoremstyle{Theorem}{3pt}{3pt}{}{}{\itshape}{---}{.5em}{}
\theoremstyle{plain}
\newtheorem{thm}{Theorem}[section]
\newtheorem{lem}{Lemma}[section]
\newtheorem{cor}{Corollary}[section]
\newtheorem{rem}{Remark}[section]

\selectlanguage{english}

\maketitle

\begin{abstract}
Recently Ivanov and Skvortsov introduced continuous Dyson-Maleev (DM) representations of supersymmetric non-linear sigma models and motivated that these representations are non-perturbatively exact. Basic to all continuous DM representations of this kind are certain identities for integrals over hermitian (super)symmetric spaces. We establish these basic identities rigorously in the non-super case.

\end{abstract}

\section{Introduction}
The Dyson-Maleev (DM) representation \cite{dys56,mal58} of spin operators in terms of bosonic creation and annihilation operators 
was originally developed and applied in the context of magnetism, more specifically in spin wave theory. In this area of research it remains a useful tool until today \cite{ivasen04}. In its simplest form the DM representation is given by
\begin{align} \label{simple_DM}
 \hat{S}_+= \hat{a}^\dagger (2S - \hat{a}^\dagger \hat{a})  \, , \quad \hat{S}_- = \hat{a} \, ,  \quad \hat{S}_z = S - \hat{a}^\dagger \hat{a} \; 
\end{align}
with the usual commutation relations $[\hat{S}_+,\hat{S}_-]=2\hat{S}_z$, $[\hat{S}_z, \hat{S}_\pm]= \pm \hat{S}_\pm$ and $\hat{a}^\dagger, \, \hat{a}$ being bosonic creation and annihilation operators. In particular $S$ denotes the spin quantum number of an $\text{SU}(2)$ representation of dimension $2S+1$. The main feature of the DM representation is the representation of spin operators through at most cubic polynomials in the bosonic operators. However, the price is that $\hat{S}_+$ fails to be the hermitian conjugate of $\hat{S}_-$.

The DM representation has also been studied intensively in the context of nuclear physics. The main objective there is to connect the study of nuclear shape vibrations to a problem of coupled oscillators. Such approaches are called boson expansions. The DM representations and generalizations thereof constitute a subclass of these boson expansion theories \cite{klein91}. 

Similar remarks can be made concerning the Holstein-Primakoff (HP) representation of spin operators \cite{hol40} which (in its simplest form) is given by
\begin{align} \label{simple_HP}
 \hat{S}_+=  \hat{a}^\dagger \sqrt{2S - \hat{a}^\dagger \hat{a}} \, , \quad \hat{S}_- =\sqrt{2S - \hat{a}^\dagger \hat{a}} \, \hat{a} \, ,  \quad \hat{S}_z = S - \hat{a}^\dagger \hat{a} \; .
\end{align}
The HP representation differs substantially from the DM representation. On the one hand $\hat{S}_+$ is the hermitian adjoint of $\hat{S}_-$ in the HP representation, which is not true in the DM representation. On the other hand the HP representation involves arbitrary powers of creation and annihilation operators whereas the DM representation involves at most cubic terms.

Using for example generalized coherent states (in the sense of Perelomov \cite{perel85}) one can change the perspective on \eqref{simple_DM} and \eqref{simple_HP} (and generalizations thereof) and regard the operators as differential operators on holomorphic functions. This approach is named generator coordinate method \cite{doba81}. In the following we refer to \eqref{simple_DM} and \eqref{simple_HP} and all generalizations thereof as the algebraic DM or HP representations.

Furthermore path integrals corresponding to generalized coherent states lead to very suggestive connections between coordinate transformations and algebraic relations such as \eqref{simple_HP}. For spin path integrals corresponding to spin coherent states $|z\rangle := \exp(z\hat{S}_+) |S, -S\rangle$ such a coordinate transformation is given by sending $z \mapsto z/\sqrt{2S-\bar{z}z}$. At the level of expectation values of spin operators $S_\alpha(z,\bar{z}):=\langle z| \hat{S}_\alpha |z \rangle/ \langle z| z \rangle$ this gives rise to 
\begin{align} 
 S_+(z,\bar{z}) \; \mapsto\;   \bar{z}\sqrt{2S - \bar{z} z } \; ,  \quad S_-(z,\bar{z})   \; \mapsto\;   \sqrt{2S - \bar{z} z }\, z \; ,  \quad S_z(z,\bar{z})   \; \mapsto\; S - \bar{z}z \; . \label{simple_HP_continuous}
\end{align}
The algebraic HP representation \eqref{simple_HP} is related to the coordinate transformation above by $a \mapsto z$ and $a^\dagger \mapsto \bar{z}$. Analogies of this kind have been studied for generalized coherent states by Mead and Papanicolaou \cite{mead83}.

Kolokolov \cite{kolok00,kolok86} established a continuous version of the algebraic DM representation in the spin case. It is obtained by making the transformation $(z,\bar{z}) \mapsto (z/(2S-\bar{z}z), \bar{z})$ which leads to 
\begin{align*}
 S_+(z,\bar{z})    \; \mapsto\;   \bar{z}(2S - \bar{z} z ) \; ,  \quad S_-(z,\bar{z})   \; \mapsto\;   z \; ,  \quad
S_z(z,\bar{z})  \; \mapsto\; S - \bar{z}z \; .
\end{align*}
In this case the transformation cannot be seen as a simple coordinate transformation since $z$ and $\bar{z}$ are transformed independently. Thus it is a nontrivial fact that such a transformation leads to an identity of the form
\begin{align} \label{spin_dm_identity}
 \int_{\mathbb{C}} \frac{dzd\bar{z}}{(1+\bar{z}z)^2} \; f(S_+, S_-, S_z) = 2S \int\limits_{\{z \in \mathbb{C}\mid \bar{z}z \leq 2S\}} dzd\bar{z} \; f\left(\bar{z}(2S - \bar{z} z ), z, S- \bar{z}z\right)
\end{align}
which is claimed to hold for holomorphic functions $f$. We refer to the right hand side of \eqref{spin_dm_identity} (and generalizations thereof) as continuous DM representation. The main attractive features of the continuous DM representation are that on the one hand the measure is flat Lebesgue measure and that on the other hand the arguments of the function $f$ are at most cubic polynomials in $z$ and $\bar{z}$. 
It is also remarkable that the two domains of integration have different topologies. On the left hand side we integrate essentially over a sphere, whereas on the right hand side we integrate over a disc.

In the context of disordered systems  Ivanov and Skvortsov \cite{ivan08} recently discussed a generalized continuous DM representation of supersymmetric non-linear sigma models of unitary type. In contrast to earlier publications \cite{gruz97,ivan06} which used such representations only at the level of perturbation theory they suggested an exact version of the continuous DM representation. However no proof was given.

Regardless of the concrete setting in which continuous DM representations are applied they rely on basic integral identities similar to \eqref{spin_dm_identity}. The purpose of this article is to establish these identities rigorously in the non-super case for the three families of hermitian symmetric spaces. In Cartan notation these are AIII, CI and DIII and in the non-linear $\sigma$ model terminology these correspond to the symmetry classes A, C and D.

The results which are stated in the first part of the article encompass the continuous DM and HP representation as distinguished members of a family of possible representations. The family of continuous representations arises from a homotopy connecting  continuous representations of HP and DM type. The essential part of the proof which is given in the second part of the article consists of a careful application of Cauchy's principle in higher dimensions. In this way we implement rigorously the ideas which were put forward in \cite{ivan08}.

\section{Results}
To state our result we have to define the setting for the three types of hermitian symmetric space it applies to. Note that each type of hermitian symmetric space (i.e. AIII, CI or DIII) gives rise to a family of corresponding compact or non-compact symmetric spaces. The symmetric spaces are introduced as quotients of certain groups. All groups that are needed are defined as fix point sets of involutions on SL$(\mathbb{C}^{p+q})$ with $p \geq q$. Using $s:= \text{Diag}(\mathbbm{1}_p, - \mathbbm{1}_q)$ the involutions are defined by
\begin{align}
&\tau(g) = (g^{\dagger})^{-1}  \;, \quad \tau'(g) = s(g^{\dagger})^{-1}s \; , \quad  
\gamma_2(g) = \Sigma_{y}  (g^t)^{-1} \Sigma_{y}  \; , \quad \gamma_1(g) = \Sigma_x  (g^t)^{-1} \Sigma_x  \; ,
\end{align}
with $\Sigma_y= \sigma_y \otimes \mathbbm{1}_N$ and $\Sigma_x = \sigma_x \otimes \mathbbm{1}_N$. For the involutions $\gamma_i$ to be defined we set $p=q=N$. The corresponding groups are summarized in table \ref{tab_involutions} and \ref{tab_groups}, e.g. 
\begin{align*}
\mathrm{SO}^*(2N)= \{ g \in \mathrm{SL}(\mathbb{C}^{2N}) \mid \tau'(g)= g = \gamma_1(g)\} \; .
\end{align*}

The last two lines of table \ref{tab_groups} contains the definitions of the complex vector spaces $W$ and $V$ that are needed to define the continuous DM representation. In the following we omit the word `continuous' and just refer to the DM representation.

\begin{thm} \label{thm_dm} Let $f$ be an analytic function on $End(\mathbb{C}^{p+q})$, $dg$ be the left invariant measure on $G/K$ ($G'/K$) and $dbd\bar{b}$ denotes the flat Lebesgue measure on $W$. Then there exists a constant $c \in \mathbb{C} \setminus \{0\}$ which does not depend on $f$ such that 

\begin{itemize}
\item[i)]  
\begin{align} \label{thm_dm_compact}
\int_{G/K} f(g s g^{-1}) dg = c \int_{\{b \in W\mid bb^{\dagger} < 1\}} f
\begin{pmatrix}
 1-2bb^{\dagger} & 2b(1 - b^{\dagger}b)^{\frac{1+t}{2}} \\
2 b^\dagger (1-bb^\dagger)^{\frac{1-t}{2}} & -1+2b^\dagger b
\end{pmatrix} db d\bar{b}
\end{align}
holds for all $t \in [0,1]$.
\item[ii)] 
\begin{align} \label{thm_dm_non_compact}
\int_{G'/K} f(g s g^{-1}) dg= c \int_{W} f
\begin{pmatrix}
 1+ 2bb^\dagger & 2b (1 + b^\dagger b)^{\frac{1+t}{2}} \\
-2 b^\dagger (1+bb^\dagger)^{\frac{1-t}{2}} & -1-2b^\dagger b
\end{pmatrix} 
 db d\bar{b}
\end{align}
holds for all $t \in [0,1]$ if the left and right hand side exists for all $t \in [0,1]$.
\end{itemize}
\end{thm}
\begin{table}[h]
\begin{center}
\begin{tabular}{c|c|c|c} 
 Cartan & AIII & CI & DIII  \\ \hline
$G$ & $\tau$ & $\tau, \; \gamma_2$ & $\tau, \; \gamma_1$ \\
$G'$& $\tau'$ &$\tau', \; \gamma_2$ & $\tau', \gamma_1 \;  $\\
$K$ & $\tau, \; \tau'$  & $\tau, \; \tau', \; \gamma_2 $ & $\tau, \; \tau' , \; \gamma_1$ \\
\end{tabular}
\end{center}
\caption{\label{tab_involutions} The non-compact group $G'$ and compact groups $G$ and $K$ are fixed by the given involutions. The first line corresponds to the nomenclature of symmetric spaces $G^{(')}/K$ introduced by Cartan.}
\end{table}
\begin{table}[h]
\begin{center}
\begin{tabular}{c|c|c|c} 
  Cartan & AIII & CI ($p=q=N$)& DIII ($p=q=N$)  \\ \hline
$G$ & SU($p+q$)  & USp($N$) & SO($2N$) \\
$G'$& SU($p,q$) & Sp($N$) & $\text{SO}^*$($2N$) \\
$K$ & S(U($p$)$ \times $U($q$)) & U($N$) & U($N$) \\
$W $& $\text{Hom}(\mathbb{C}^{p}, \mathbb{C}^q) $ &$\{Z \in \text{End}(\mathbb{C}^N) \mid Z^t=Z\}$ & $\{ Z \in \text{End}(\mathbb{C}^N) \mid Z^t=-Z \}$ \\
$V $& $\text{Hom}(\mathbb{C}^{q}, \mathbb{C}^p) $ &$\{Z \in \text{End}(\mathbb{C}^N) \mid Z^t=Z\}$ & $\{ Z \in \text{End}(\mathbb{C}^N) \mid Z^t=-Z \}$ 
\end{tabular}
\end{center}
\caption{\label{tab_groups}$G/K$ is the compact and $G'/K$ the non-compact symmetric space. $W$ and $V$ are needed as parameter spaces for the corresponding symmetric spaces.}
\end{table}

\begin{rem}
 For $t=0$ the right hand side constitutes the HP representation and for $t=1$ the DM representation.
\end{rem}

\begin{rem}
The two advantages of the DM representation are clearly visible. Firstly the matrix entries of $gsg^{-1}$ are at most cubic in $b$ and secondly the measure is flat. Both features combined represent a substantial simplification for concrete calculations.
\end{rem}

\begin{rem}
 The version of the DM and HP representations stated here differs by a trivial rescaling $b \mapsto 2b$ from the one given in \cite{ivan08} since we integrate over unit disks.
\end{rem}

\section{Proof}

In the following we give a proof of theorem \ref{thm_dm}. Notice that we handle all cases at once. Using the involution $\theta(g)=sgs^{-1}$ on SL$(\mathbb{C})$ we can define for each symmetry class the quotient $G^{\mathbb{C}}/K^{\mathbb{C}}$ (see table \ref{Complex_groups}). 
\begin{table}[!ht]
\begin{center}
\begin{tabular}{c|c|c|c} 
  Cartan& AIII& CI & DIII  \\ \hline
$G^{\mathbb{C}}$ & - & $\gamma_2$ & $\gamma_1$ \\
$K^{\mathbb{C}}$ & $\theta$  & $\theta, \; \gamma_2 $ & $\theta , \; \gamma_1$ \\
\end{tabular}
\end{center}
\caption{\label{Complex_groups} The groups $G^{\mathbb{C}}$ and $K^{\mathbb{C}}$ are fixed by the given involutions.}
\end{table}
In the proof we view $G^{(')}/K$ as domain in $G^{\mathbb{C}}/K^{\mathbb{C}}$. This subspace is then deformed by the Cauchy principle into another domain which is called the Dyson-Maleev domain. For theorem \ref{thm_dm} the choice of parametrization is obviously important. To set the stage we fix a concrete realization of $G^{\mathbb{C}}/K^{\mathbb{C}}$:
\begin{align*}
 \vartheta : G^{\mathbb{C}}/K^{\mathbb{C}} &\rightarrow \mathrm{End}(\mathbb{C}^{p+q}) \\
[g] &\mapsto g s g^{-1} \; .
\end{align*}
A common choice of coordinates on a certain patch $\mathcal{U} \subset G^{\mathbb{C}}/K^{\mathbb{C}}$ is given by (see \cite{berc97})
\begin{align*}
  \varphi: \mathcal{U} &\rightarrow W \times V \\
\left[\begin{pmatrix} A & B \\ C & D \end{pmatrix}\right] &\mapsto ( B D^{-1} , CA^{-1}) \; ,
\end{align*}
where the patch $\mathcal{U}$ covers the region where the blocks $A$ and $D$ are invertible. The parametrization of the symmetric space in its concrete realization given by $\vartheta$ is $Q:=\vartheta \circ \varphi^{-1}$. A short computation shows that
\begin{align*}
Q: \varphi(\mathcal{U}) &\rightarrow \mathrm{End}(\mathbb{C}^{p+q})  \\
(Z, \tilde{Z}) &\mapsto \begin{pmatrix}
(1+Z\tilde{Z})(1-Z\tilde{Z})^{-1} & -2 Z(1-\tilde{Z}Z)^{-1} \\
2 \tilde{Z} (1-Z\tilde{Z})^{-1} & -(1+\tilde{Z}Z)(1-\tilde{Z}Z)^{-1} 
\end{pmatrix} \; .
\end{align*}
We want to view $Q$ as a holomorphic function on a certain subset of $W\times V$. 
In order to explain this let us consider functions of the form $G(x):=(1+X)^t:=\exp (t\ln (1+X))$, where $X$ is a complex square matrix and $t \in \mathbb{R}$ is fixed. A series expansion around $X=0$ yields a well defined holomorphic function in $X$ which can be analytically continued. Since we apply the mapping without approaching the branch cut of the logarithm it is not necessary to specify the domain of definition explicitly. Similar remarks apply to the mapping
\begin{align}
R(b, \tilde{b}) = \left(-b (1+\tilde{b}b)^{-\frac{1}{2}},-\tilde{b} (1+b\tilde{b})^{-\frac{1}{2}}\right)
\end{align}
for $b\in W$ and $\tilde{b} \in V$. We obtain
\begin{align} \label{Q_circ_R}
 (Q \circ R)(b,\tilde{b})= 
\begin{pmatrix}
 1+ 2b\tilde{b} & 2b (1 + \tilde{b}b)^{\frac{1}{2}} \\
-2 \tilde{b} (1+b\tilde{b})^{\frac{1}{2}} & -1-2\tilde{b}b
\end{pmatrix} \; .
\end{align}
In the coordinate patch defined by $Q$ the invariant two form $\omega$ on $G^{\mathbb{C}}/K^{\mathbb{C}}$ is given by (see \cite{berc97})
\begin{align*}
 Q^* \omega &= \Tr\left((1-Z\tilde{Z})^{-1} dZ \wedge (1-\tilde{Z}Z)^{-1}d\tilde{Z}\right) \\
&= d \Tr \left(Z (1-\tilde{Z}Z)^{-1}d\tilde{Z} \right)  \; .
\end{align*}
Notice that the last equality is only valid in the coordinate patch defined by $Q$ and not globally since $\omega$ is not exact. Now we can use the mapping $R$ to introduce normal coordinates
\begin{align}
 (Q\circ R )^* \omega =  d \Tr \left(b d\tilde{b}\right) + d \Tr\left(b \tilde{b} (1+b\tilde{b})^{\frac{1}{2}} d(1+b\tilde{b})^{-\frac{1}{2}}\right)= \Tr \left(db \wedge d\tilde{b}\right)\; , \label{flat_two_form}
\end{align}
and hence the invariant holomorphic volume form in the coordinate patch defined by $Q \circ R$ is given by
\begin{align*}
\Omega := \bigwedge^{\mathrm{dim}_{\mathbb{C}}W} \Tr(db \wedge d\tilde{b}) \; .
\end{align*}

\subsection{Non-compact case}
The non-compact symmetric space G$'$/K is parametrized by $Q(Z, Z^{\dagger})$ with $Z \in W$ and $ZZ^\dagger < 1$ (See \cite{berc97}). Alternatively we can use that the mapping
\begin{align*}
W &\rightarrow \{Z \in W\mid ZZ^{\dagger}< 1\}  \\
Z &\mapsto Z(1+ Z^{\dagger}Z)^{-\frac{1}{2}}
\end{align*}
is a diffeomorphism to obtain $(Q \circ R)(b, b^\dagger)$ (with $b \in W$) as a different parametrization of G$'$/K. In view of equation \eqref{Q_circ_R} and \eqref{flat_two_form} we know that up to a normalization factor the left hand side of \eqref{thm_dm_non_compact} equals:
\begin{align} \label{non_compact_darboux}
\int\limits_{\substack{W \rightarrow W \times V \\ b \;\mapsto (b, b^\dagger)}} f\circ Q \circ R \; \Omega \; .
\end{align}
Consider the homotopy given by
\begin{align*}
H_+^r: [0,1] \times \{b \in W\mid bb^\dagger \leq r\} &\rightarrow W \times V \\
(t, b)&\mapsto (b (1+ b^\dagger b)^{t/2},b^\dagger (1+ b b^\dagger)^{-t/2} ) \; .
\end{align*}
In contrast to the compact case, which will be discussed below, we have a global chart for G$'$/K. The restriction of the homotopy $H_+^r$ to a matrix ball with radius $r$ is necessary for a clean discussion of potential boundary terms. 

Note that since $f$ is a holomorphic function which is composed with holomorphic functions of $b$ and $\tilde{b}$ the integrand of \eqref{non_compact_darboux} is a closed form, i.e. 
\begin{align}\label{hol_nc}
 d (f \circ Q \circ R \; \Omega ) =0 \; .
\end{align}
Thus we may use the Cauchy principle (or Stokes' theorem) to deform the domain of integration using the homotopy $H_+^r$:
\begin{align*}
 0 \overset{\eqref{hol_nc}}{=} \int_{H_+^r} d (f\circ Q\circ R \; \Omega) = &\int_{H_+^r|_{t=1}} f\circ Q \circ R
\; \Omega  - \int_{H_+^r|_{t=0}} f\circ Q \circ R \; \Omega \; .\\
 & +\int_{H_+^r|_{[0,1]\times \partial \{b \in W| bb^\dagger \leq r \}}}  f\circ Q\circ R_- \; \Omega  \;.
\end{align*}
All integrands in the above equation are well behaved due to the assumption that the the right hand side in \eqref{thm_dm_non_compact} exists for all $t \in [0,1]$. Provided that the last term on the right hand side vanishes for all positive $r$ we get equation \eqref{thm_dm_non_compact} in the limit  $r \rightarrow \infty$.
To obtain the desired result it is thus sufficient to prove the following lemma:
\begin{lem} \label{non_compact_lem} For all $r \in \mathbb{R}_+$ 
\begin{align*}
\left(H_+^r|_{[0,1]\times \partial \{b \in W\mid bb^\dagger \leq r \}} \right)^* \Omega =0 \; .
\end{align*}
\end{lem}
This lemma contains the crucial part of the proof. Its proof is given in the last section. In the next section we deal with the compact case which is slightly more complicated but also crucially relies on a lemma similar to the one above.

\subsection{Compact version of DM}
The compact symmetric space G/K is parametrized up to a set of measure zero by $Q(Z, -Z^\dagger )$ with $Z \in W$ (see \cite{berc97}). We define $R_-(b, \tilde{b}):= R(b, -\tilde{b}) $ and use that the mapping
\begin{align*}
\{Z \in W\mid ZZ^{\dagger}< 1\} &\rightarrow W  \\
Z &\mapsto Z(1- Z^{\dagger}Z)^{-\frac{1}{2}}
\end{align*}
is a diffeomorphism (in fact it is the inverse of the analogous mapping in the non-compact case) to obtain $(Q\circ R_-)(b, b^\dagger)$ with $b \in W$ as a different parametrization of G/K. Making use of equation \eqref{Q_circ_R} and \eqref{flat_two_form} we see that up to a normalization factor the left hand side of \eqref{thm_dm_compact} equals
\begin{align*}
\int\limits_{\substack{\{b \in W \mid bb^\dagger < 1\} \rightarrow W \times V \\ b \;\mapsto (b, b^\dagger)}} f\circ Q \circ R_- \; \Omega \; .
\end{align*}
Let us introduce the homotopy
\begin{align*}  
H_-^r: [0,1] \times \{b \in W\mid bb^\dagger \leq r\} &\rightarrow W \times V \notag \\
(t, b)&\mapsto (b (1- b^\dagger b)^{t/2},b^\dagger (1- b b^\dagger)^{-t/2} ) \; . 
\end{align*}
We have to demand $0 <r <1$ in the definition above to allow for a discussion of possible boundary contributions. This is not only necessary because the coordinate patch we use does not cover the compact domain G/K completely but is also dictated by the fact that $H_-^1$ is singular outside the coordinate patch. The latter can be seen for $t=1$ in the lower left block of
\begin{align} \label{Q_deformation}
 (Q \circ R_- \circ H_-^r)(b) = 
\begin{pmatrix}
 1- 2bb^{\dagger} & 2b (1 - b^{\dagger}b)^{\frac{1+t}{2}} \\
2 b^{\dagger} (1-bb^{\dagger})^{\frac{1-t}{2}} & -1+2b^{\dagger}b
\end{pmatrix} \; .
\end{align}
On the domain where $Q\circ R_-$ is a holomorphic function we have 
\begin{align} \label{hol_c}
d(f\circ Q\circ R_- \; \Omega)=0 \; . 
\end{align}
Application of Cauchy's principle using the regularized homotopy $H_-^r$ with $0<r<1$ leads to 
\begin{align}
 0 \overset{\eqref{hol_c}}{=} \int_{H_-^r} d\left(f\circ Q\circ R_- \;  \Omega\right)= &\int_{H_-^r|_{t=1}}  f\circ Q\circ R_- \; \Omega 
-\int_{H_-^r|_{t=0}}  f\circ Q\circ R_- \; \Omega \notag \\
 & +\int_{H_-^r|_{[0,1]\times \partial \{b \in W\mid bb^\dagger \leq r \}}}  f\circ Q\circ R_- \; \Omega  \;. \label{Cauchy_compact}
\end{align}

The following lemma asserts that the last term in the above equation vanishes. 

\begin{lem} \label{compact_lem} For all $r\in ]0,1[$
\begin{align*}
\left(H_-^r|_{[0,1]\times \partial \{b \in W\mid bb^\dagger \leq r \}} \right)^* \Omega =0 \; .
\end{align*}
\end{lem}
This lemma which contains the crucial part of the proof of theorem \ref{thm_dm} will be proved below together with lemma \ref{non_compact_lem}. 

In view of \eqref{Cauchy_compact} we conclude that for all $r\in ]0,1[$
\begin{align} \label{int_eq_delta}
 \int_{H_-^r|_{t=1}}  f\circ Q\circ R_- \; \Omega  = \int_{H_-^r|_{t=0}}  f\circ Q\circ R_- \; \Omega \; .
\end{align}
Translating \eqref{int_eq_delta} to Lebesgue integrals yields
\begin{align*}
  \int_{W} &\chi_r(b) \; f
\begin{pmatrix}
 1-2bb^{\dagger} & 2b(1 - b^{\dagger}b)^{\frac{1}{2}} \\
-2 b^\dagger (1-bb^\dagger)^{\frac{1}{2}} & -1+2b^\dagger b
\end{pmatrix} db d\bar{b}\\ &= \int_{W} \chi_r(b) \; f
\begin{pmatrix}
 1-2bb^{\dagger} & 2b(1 - b^{\dagger}b) \\
-2 b^\dagger  & -1+2b^\dagger b
\end{pmatrix} db d\bar{b} \; ,
\end{align*}
where $\chi_r$ denotes the characteristic function of $\{b \in W\mid bb^{\dagger} \leq r\}$. Since $\chi_r \; f\circ Q \circ R_- \circ H_-^1$ is a bounded function which converges pointwise to $\chi_1 \; f\circ Q \circ R_- \circ H_-^1$  we can apply Lebesgues theorem of dominated convergence to obtain in the limit $r \rightarrow 1$
\begin{align*}
  \int_{\{b \in W\mid bb^{\dagger} \leq 1\}} &f
\begin{pmatrix}
 1-2bb^{\dagger} & 2b(1 - b^{\dagger}b)^{\frac{1}{2}} \\
-2 b^\dagger (1-bb^\dagger)^{\frac{1}{2}} & -1+2b^\dagger b
\end{pmatrix} db d\bar{b}\\ &= \int_{\{b \in W\mid bb^{\dagger} \leq 1\}} f
\begin{pmatrix}
 1-2bb^{\dagger} & 2b(1 - b^{\dagger}b) \\
-2 b^\dagger  & -1+2b^\dagger b
\end{pmatrix} db d\bar{b} \; .
\end{align*}
In the above argument we used the homotopy to deform from $t=0$ to $t=1$. Of course we can stop the deformation at any value $0\leq t\leq 1$ and thus we obtain equation \eqref{thm_dm_compact}.

\subsection{Proof of lemma \ref{non_compact_lem} and \ref{compact_lem}}
In the following we prove both lemmas in parallel. First we observe that for each $b\in W$ there exists a singular value decomposition of the form $b = u D v^\dagger$, where $u, v$ are appropriate unitary matrices and $D$ is a diagonal matrix (note that in the AIII case the matrix b is allowed to be rectangular). To make this more precise and to  obtain a corresponding parametrization of $W$ we apply some results from Lie theory \cite{knapp05}. Let $G'$ be a semisimple Lie group with Lie algebra $\mathfrak{g}$, whose Cartan decomposition is given by $\mathfrak{k}\oplus \mathfrak{p}$. Furthermore let $K$ denote the corresponding compact subgroup of $G'$, $\mathfrak{a}$ a maximal abelian subalgebra of $\mathfrak{p}$ and $Z_K(\mathfrak{a})$ the centralizer of $\mathfrak{a}$ in $K$. If $\mathfrak{a}_+$ denotes a positive Weyl chamber, then
\begin{align}
 &K/Z_{K}(\mathfrak{a}) \times \mathfrak{a}_+ \rightarrow \mathfrak{p} \notag \\
&([k], H) \mapsto k H k^{-1} \label{diagonalize_p}
\end{align}
is a suitable parametrization of $\mathfrak{p}$. It is a fact that the mapping above is a diffeomorphism onto its image if it is restricted to the interior of $\mathfrak{a}_+$. In the concrete cases we discuss $\mathfrak{p}$ is closely related to $W$ and it is easily possible to use \eqref{diagonalize_p} to obtain a diagonalization result for $W$.

For definiteness we consider the AIII case with $p\geq q$, then $G'=SU(p,q)$ and $K=S(U(p)\times U(q))$ and 
\begin{align} \label{p_and_w}
\mathfrak{p} = \left\{\left(\begin{smallmatrix}0 & Z \\ Z^{\dagger} & 0  \end{smallmatrix}\right) \mid Z \in W \right\} \; .
\end{align}
As maximal abelian subalgebra we use
\begin{align*}
\mathfrak{a}= \left\{ 
\begin{pmatrix}
0 & D \\
D^{\dagger} & 0
\end{pmatrix}
\bigg|
D = 
\begin{pmatrix}
 D_1 \\
 & \ddots \\
 & & D_q \\
 & 0 & \\
\end{pmatrix} , D_i \in \mathbb{R} \right\} \; .
\end{align*}
Then we define $f_i: \mathfrak{a} \rightarrow \mathbb{R}$ to be $f_i\left(\begin{smallmatrix} 0 &D \\ D^{\dagger} & 0 \end{smallmatrix}\right) = D_i$. Then the restricted roots, i.e. the simultaneous eigenvalues of the commutator action of $\mathfrak{a}$ on $\mathfrak{k}\oplus \mathfrak{p}$ for $1\leq k < l \leq q$ are given by
\begin{align*}
 \pm f_k, \; \pm (f_k - f_l), \;\pm (f_k + f_l) \; .
\end{align*}
Note that the roots $f_k$ appear only for $p\neq q$. Let $\Sigma_+$ denote a set of positive roots and fix the corresponding 
positive Weyl chamber to be
\begin{align*}
 \mathfrak{a}_+ = \{H \in \mathfrak{a}\mid  \alpha(H)\geq 0  \, , \forall \alpha \in \Sigma_+ \} \; .
\end{align*}
For $k = \left(\begin{smallmatrix}u & 0 \\ 0 & v \end{smallmatrix}\right) \in K$ and $H=\left(\begin{smallmatrix}0 & D \\ D^{\dagger} & 0 \end{smallmatrix}\right) \in \mathfrak{a}$ we have 
$k H k^{-1} =
\left(\begin{smallmatrix}
 0 & u D v^{\dagger} \\
vD^{\dagger}u^{\dagger}& 0
\end{smallmatrix}\right)$.
Thus referring to \eqref{diagonalize_p} and \eqref{p_and_w} we have the following reparametrization of $W$
\begin{align*}
 \Phi: \,&K/Z_{K}(\mathfrak{a}) \times \mathfrak{a}_+ \rightarrow W \\ 
&\left( \left[ \begin{pmatrix}
  u & 0 \\
0 & v
 \end{pmatrix} \right], \begin{pmatrix} 0 & D \\ D^{\dagger} & 0 \end{pmatrix}\right) 
\mapsto 
uDv^{\dagger} \; .
\end{align*}
Notice that the homotopy $H_\pm^r$ is compatible with $\Phi$ in the sense that
\begin{align*} 
 H_\pm^r(t, u D v^\dagger)= \left(u D(1\pm DD^\dagger)^{t/2} v, v D^\dagger(1\pm DD^\dagger)^{-t/2} u^\dagger \right) \; . 
\end{align*}
Next we compute the pull back of $\Tr(db\wedge d\tilde{b})$ with $H_\pm^r\circ (id_{[0,1]}, \Phi)$. For this it will be convenient to use the following abbreviations $D'\equiv D(1\pm DD^\dagger)^{t/2}$, $\tilde{D}'\equiv D^\dagger(1\pm DD^\dagger)^{-t/2}$, 
\begin{align*}
 H' \equiv \begin{pmatrix}
  0 & D' \\
\tilde{D}' & 0
 \end{pmatrix} 
\text{ and } 
[k] \equiv \Big[\begin{pmatrix}
       u & 0 \\
	0 & v
      \end{pmatrix} \Big]\in K / Z_{K}(\mathfrak{a}) \; .
\end{align*}
Then we have
\begin{align}
&\big(H_\pm^r\circ (id_{[0,1]}, \Phi) \big)^*\Tr(db \wedge d\tilde{b}) = \frac{1}{2} \Tr\left( s d(kH' k^{-1})\wedge  d(k H' k^{-1}) \right) \notag \\
&=\frac{1}{2} \Tr(sdH' \wedge dH') + \Tr( s k^{-1}dk \wedge d(H'^2)) + \frac{1}{2} \Tr( k^{-1}dk \wedge [k^{-1}dk, sH'^2]) \label{eval_two_form}
\end{align}
The first two terms in \eqref{eval_two_form} contain only diagonal terms and evaluate to 
\begin{align}
  \sum_{i=1}^q &dD_i' \wedge d\tilde{D}_i' - (v^\dagger dv - u^\dagger du)_{ii}  \wedge d(D_i' \tilde{D}_i')  \notag \\ &= \sum_{i=1}^q dD_i \wedge D_i\big(2 (v^\dagger dv - u^\dagger du)_{ii} - \ln(1\pm D_i^2) dt  \big)\; , \label{first_two}
\end{align}
and the last term in \eqref{eval_two_form} equates to 
\begin{align}
\sum_{1\leq k < l \leq p} &\left((D'\tilde{D}')_k-(D'\tilde{D}_l')\right) (u^\dagger du)_{kl} \wedge (u^\dagger du)_{lk} \notag\\
&-\sum_{1\leq k < l \leq q} \left((\tilde{D}'D')_k-(\tilde{D}'D')_l\right) (v^\dagger dv)_{kl} \wedge (v^\dagger dv)_{lk} \;. \label{last}
\end{align}
The pull back of $\Omega$ can be computed by taking the $\dim_{\mathbb{C}} W$ fold wedge product of the pull back of \eqref{eval_two_form} and the only nontrivial contribution to this wedge product is proportional to the $q$-fold wedge product of  \eqref{first_two} wedged with the $(\dim_{\mathbb{C}} W - q)$-fold wedge product of \eqref{last}. Hence we obtain
\begin{align}
(H_\pm^r \circ (id_{[0,1]},\Phi))^*\Omega =  \bigwedge_{i=1}^q dD_i \bigwedge \left(\dots \right)
\; , \label{AIII_jacobian} 
\end{align}
where the dots represent terms whose precise form is of no importance for the following argument. 
If $H_\pm^r$ is  restricted to $[0,1]\times \partial \{b \in W\mid bb^\dagger \leq r \}$, i.e. $\Phi$ is restricted to $K/Z_K(\mathfrak{a})\times \partial \{H \in \mathfrak{a}_+ \mid H^2 \leq r\}$ the one forms $dD_i$ are no longer independent and hence \eqref{AIII_jacobian} equates to zero.

Concerning the other symmetry classes: 
For symmetric spaces of type CI and DIII we specify the abelian subalgebra $\mathfrak{a}\subset \mathfrak{p}$ in the appendix. Using the explicit forms of $\mathfrak{a}$ it can be easily checked that the reasoning leading to \eqref{first_two}, \eqref{last} and \eqref{AIII_jacobian} is the same.

{\bf Acknowledgement.} This work was supported by the Deutsche Forschungsgemeinschaft (SFB/TR 12). I thank M. Zirnbauer and P. Heinzner for helpful discussions.

\section{Appendix}
We briefly list the information needed to complete the proof for symmetry classes CI and DIII. We use the notation introduced in tables \ref{tab_involutions} and \ref{tab_groups}.

\subsection{Type CI}
In addition to the objects already defined in the tables we introduce

\begin{align*}
 \mathfrak{a}= \left\{ 
\begin{pmatrix}
0 & D \\
D^{\dagger} & 0
\end{pmatrix}
\bigg|
D = 
\begin{pmatrix}
D_1 \\
 & \ddots \\
 & & D_N \\
\end{pmatrix} , D_i \in \mathbb{R} \right\}
\end{align*}
as a maximal abelian subalgebra of
\begin{align*}
\mathfrak{p} = \left\{ 
\begin{pmatrix}
 0 & Z\\
Z^{\dagger}& 0
\end{pmatrix}
\bigg|
Z^t=Z \in \mathbb{C}^{N\times N}
\right\} \; .
\end{align*}
Furthermore note that 
\begin{align*}
 [k] = \left[\begin{pmatrix}
              u & 0 \\
	      0 & u^\dagger
             \end{pmatrix}
 \right] \in K / Z_{K}(\mathfrak{a}) \; .
\end{align*}
Thus setting $v = u^\dagger$ and $q=N$ we obtain the analogue of \eqref{eval_two_form}, \eqref{first_two} and \eqref{last}. In particular the pull back of $\Omega$ is also of the form \eqref{AIII_jacobian}.

\subsection{Type DIII}
We choose 
\begin{align*}
 \mathfrak{a}= \left\{ 
\begin{pmatrix}
0 & D \\
D^{\dagger} & 0
\end{pmatrix}
\bigg|
D = 
\begin{pmatrix}
 d_1 i \sigma_y \\
 & \ddots \\
 & & d_{N/2} i \sigma_y \\ 
\end{pmatrix} , d_i \in \mathbb{R} \right\}
\end{align*}
as maximal abelian subalgebra of
\begin{align*}
 \mathfrak{p} = \left\{
\begin{pmatrix}
 0 & Z \\
Z^{\dagger} &0
\end{pmatrix}
\bigg|
-Z^{t}= Z \in \mathbb{C}^{N\times N}
 \right\} \; .
\end{align*}
Note that we have to add a row of zeros to $D$ if $N$ is odd and that 
\begin{align*}
 D^{\dagger}D'= 
\begin{pmatrix}
d_1 d_1' \sigma_0& & \\
& \ddots & \\
& & d_N d_N'\sigma_0
\end{pmatrix}
\end{align*}
The compact group $K$ is the same as in the CI case and we have
\begin{align*}
 [k] = \left[\begin{pmatrix}
              u & 0 \\
	      0 & u^\dagger
             \end{pmatrix}
 \right] \in K / Z_{K}(\mathfrak{a}) \; .
\end{align*}
As in the case of CI we set $v = u^\dagger$ and $q=N$ which also leads to \eqref{AIII_jacobian}.

\end{document}